%% file: paper.tex
\documentclass[twocolumn,showpacs,aps,prl,superscriptaddress]{revtex4b5}

\usepackage{graphicx}
\usepackage{dcolumn}
\usepackage{amsmath}
\usepackage{epsfig}

\input pubboard/babarsym

\newcommand{\BABARPubYear}    {01}
\newcommand{\BABARPubNumber}  {01}

\newcommand{\SLACPubNumber} {8777}
\newcommand{\LANLNumber} {0102030}

\def\figurebox#1#2#3{%
    \def\arg{#3}%
    \ifx\arg\empty
    {\hfill\vbox{\hsize#2\hrule\hbox to #2{\vrule\hfill\vbox to #1{\hsize#2\vfill}\vrule}\hrule}\hfill}%
    \else
    {\hfill\epsfbox{#3}\hfill}%
    \fi}

\long\def\inst#1{\par\nobreak\kern 4pt\nobreak
    {\it #1}\par\vskip 10pt plus 3pt minus 3pt}

\begin{document}

\preprint{\babar-PUB-\BABARPubYear/\BABARPubNumber} 
\preprint{SLAC-PUB-\SLACPubNumber} 

\begin{flushleft}
\babar-PUB-\BABARPubYear/\BABARPubNumber\\
SLAC-PUB-\SLACPubNumber\\
hep-ex/\LANLNumber\\[20mm]
\end{flushleft}

\title{
\vskip 10mm
{\large \bf
 Measurement of {\boldmath \CP}-Violating Asymmetries in
{\boldmath \Bz} Decays to {\boldmath \CP}\ Eigenstates} 
\begin{center} 
\vskip 10mm
The \babar\ Collaboration
\end{center}
}

\input authors

\date{February 18, 2001}

\begin{abstract}
We present measurements of time-dependent \CP-violating asymmetries in neutral
$B$ decays to several \CP eigenstates. The measurement uses 
a data sample of 23 million $\FourS\to B\Bbar$ decays collected
by the \babar\ detector at the \pep2\ asymmetric \BF\ at SLAC.
In this sample, we find events where one neutral $B$ meson is
fully reconstructed in a \CP eigenstate containing charmonium and
the flavor of the other neutral $B$ meson is determined from its decay products.
The amplitude of the \CP-violating asymmetry, which in the Standard Model is 
proportional to \stwob, is derived from the decay time distributions in such events.
The result is $\stwob=0.34 \pm 0.20\ {\rm (stat)} \pm 0.05\ {\rm (syst)}$.
\end{abstract}


\maketitle
\par
\CP-violating asymmetries in the time distributions for decays
of \Bz and \Bzb mesons provide a direct test of the Standard Model 
of electroweak interactions~\cite{BCP}.
For the neutral $B$ decay modes reported here, corrections to \CP-violating effects 
from strong interactions are absent, in contrast to the \KL\ modes in
which \CP\ violation was discovered~\cite{KLCP}.
\par
Using a data sample of 23 million $B\Bbar$ pairs recorded
at the $\FourS$ resonance by the \babar\ detector at the PEP-II
asymmetric-energy \epem collider at the Stanford Linear
Accelerator Center,
we have fully reconstructed a sample $B_{\CP}$ of neutral $B$ mesons decaying
to the \CP eigenstates $\jpsi\KS$, $\psitwos\KS$ and $\jpsi\KL$.
We examine each of the events in this sample for
evidence that the other neutral $B$ meson
decayed as a \Bz or a \Bzb, designated as a \Bz\ or \Bzb\ flavor tag.
The final $B_{CP}$ sample contains about 360 signal events.
\par
When the \FourS\ decays, the $P$-wave \BB\ state evolves coherently 
until one of the mesons decays. In
one of four time-order and flavor configurations, 
if the tagging meson $B_{\rm tag}$ decays
first, and as a \Bz, the other meson must be a \Bzb\ at that
same time $t_{\rm tag}$. It then evolves independently, and can decay into a \CP
eigenstate $B_{CP}$ at a later time $t_{CP}$. The time between the two decays
$\deltat=t_{CP}-t_{\rm tag}$ is a signed quantity made measurable by producing the $\FourS$ with
a boost $\beta\gamma=0.56$ along the collision ($z$) axis, with nominal
energies of 9.0 and 3.1\gev\ for the electron and positron beams.
The measured distance $\deltaz\approx\beta\gamma c\deltat$ between the
two decay vertices provides a good estimate of the corresponding time interval \deltat; 
the average value of $|\deltaz|$ is 
$\beta\gamma c\tau_{\Bz}\approx 250\mum$.  
\par
The decay-time distribution for events with a \Bz or a \Bzb tag can be expressed
in terms of a complex parameter $\lambda$ that depends on both
\BzBzb mixing and on the amplitudes describing \Bzb and
\Bz decay to a common final state $f$~\cite{lambda}.
The distribution
${\rm f}_+({\rm f}_-)$ of the decay rate when the
tagging meson is a $\Bz (\Bzb)$ is given by
\begin{eqnarray}
{\rm f}_\pm(\, \deltat) = {\frac{{\rm e}^{{- \left| \deltat \right|}/\tau_{\Bz} }}{2\tau_{\Bz}
(1+|\lambda|^2) }}  \times  \left[ \ {\frac{1 + |\lambda|^2}{2}} \hbox to 2cm{}
\right. \nonumber \\ 
\left. 
\pm {\ \mathop{\cal I\mkern -2.0mu\mit m}}
\lambda  \sin{( \Delta m_{B^0}  \deltat )} 
\mp { \frac{1  - |\lambda|^2 } {2} }  
  \cos{( \Delta m_{B^0}  \deltat) }   \right],
\label{eq:timedist}
\end{eqnarray}
where $\tau_{\Bz}$ is the \Bz lifetime and $\Delta m_{B^0}$ is the mass difference determined
from \BzBzb mixing~\cite{PDG2000}, and where
the lifetime difference between neutral \B\ mass eigenstates is assumed to be negligible. 
The first oscillatory term in Eq.~\ref{eq:timedist} is due to interference between 
direct decay and decay after mixing. 
A difference between the \Bz and \Bzb distributions or
a \deltat asymmetry for either tag is evidence for \CP violation.
\par
If all amplitudes contributing to $B^0\to f$ have
the same weak phase, a condition satisfied in the Standard Model for  
charmonium-containing $b\to\ccbar s$ decays, then $|\lambda|=1$. 
For these \CP eigenstates the Standard Model predicts
$\lambda=\eta_f e^{-2i\beta}$, where $\eta_f$ is the \CP eigenvalue of
the state $f$ and
$\beta = \arg \left[\, -V_{\rm cd}^{}V_{\rm cb}^* / V_{\rm td}^{}V_{\rm tb}^*\, \right]$
is an angle of the Unitarity Triangle of the three-generation 
Cabibbo-Kobayashi-Maskawa (CKM) matrix~\cite{CKM}.
Thus, the time-dependent \CP-violating asymmetry is
\begin{eqnarray}
A_{\CP}(\deltat) &\equiv&  \frac{ {\rm f}_+(\deltat)  -  {\rm f}_-(\deltat) }
{ {\rm f}_+(\deltat) + {\rm f}_-(\deltat) } \nonumber \\%
&=& -\eta_f \stwob \sin{ (\Delta m_{B^0} \, \deltat )} , 
\label{eq:asymmetry}
\end{eqnarray}
where $\eta_f=-1$ for $\jpsi\KS$ and $\psitwos\KS$ and
$+1$ for $\jpsi\KL$. 
\par
A measurement of $A_{\CP}$ requires determination of
the experimental \deltat resolution and  
the fraction of events in which the tag assignment is incorrect. 
A mistag fraction \mistag reduces the observed
asymmetry by a factor $(1-2\mistag)$.
\par
Several samples of fully reconstructed \Bz mesons are used in this measurement.
The $B_{\CP}$ sample contains  
candidates reconstructed in the \CP eigenstates
$\jpsi\KS\ (\KS \to \pipi, \ppz)$,  
$\psitwos\KS\ (\KS \to \pipi)$ 
and $\jpsi\KL$.   
The $\jpsi$ and $\psitwos$ mesons are reconstructed through their decays to \epem
and \mumu; the $\psitwos$ is also reconstructed through its decay to $\jpsi\pipi$.
A sample of 
$B$ decays $B_{\rm flav}$~\cite{conjugates} 
used in the determination of the mistag fractions 
and \deltat resolution functions
consists of the channels $D^{(*)-}h^+(h^+=\pi^+,\rho^+,a_1^+$) and 
$\jpsi K^{*0}\ (K^{*0}\to K^+\pi^-)$.
A control sample of
charged $B$ mesons decaying to the final states $\jpsi K^{(*)+}$, 
$\psitwos K^+$  
and $D^{(*)0}\pi^+$ is used for validation studies.
\par
A description of the \babar\ detector can be found in Ref.~\cite{BABARNIM}.
Charged particles
are detected and their momenta measured by a combination of
a silicon vertex tracker (SVT) consisting of
five double-sided layers and a central drift chamber (DCH),
in a 1.5-T solenoidal field.
The average vertex resolution in the $z$ direction 
is 70\mum\ for a
fully reconstructed $B$ meson.
We identify leptons and hadrons
with measurements from all detector systems,
including the energy loss (\dedx) in the DCH and SVT. Electrons
and photons are identified by a CsI electromagnetic calorimeter
(EMC). Muons are identified in the instrumented flux return (IFR).
A Cherenkov ring imaging detector
(DIRC) covering the central region, together with the \dedx\ information,
provides $K$-$\pi$ separation of at
least three standard deviations for
$B$ decay products with momentum greater than 250\mevc in the laboratory.
\par
We select events with a minimum of three reconstructed charged tracks,
each having a laboratory polar angle between 0.41 and 2.54 rad and impact parameter in
the plane transverse to the beam less than 1.5\cm\ from the beamline.
The event must have a total measured energy in the laboratory greater than 4.5\gev 
within the fiducial regions for charged tracks and neutral clusters.
To help reject continuum background, the second Fox-Wolfram moment~\cite{fox}
must be less than 0.5.
\par
An electron candidate must have
a ratio of calorimeter
energy to track momentum, an EMC cluster shape, 
a DCH \dedx\ and a DIRC
Cherenkov angle (if available) consistent with an electron.
\par
A muon candidate must satisfy requirements on the measured and
expected number of interaction lengths 
penetrated, the position match
between the extrapolated DCH track and IFR hits, and the average
and spread of the number of IFR hits per layer.
\par
A track is identified as a kaon candidate by means of a neural network that 
uses \dedx\ measurements in the DCH and
SVT, and comparison of the observed pattern of detected photons in the DIRC with that
expected for kaon and pion hypotheses.
\par
Candidates for $\jpsi \to \ell^+\ell^-$ must have at least one 
decay product identified as a lepton (electron or muon) candidate or, 
if outside the calorimeter acceptance, must have 
DCH \dedx\ information consistent with the electron hypothesis.
Tracks in which the electron has radiated are combined with bremsstrahlung photons, reconstructed as
clusters with more than 30\mev\ lying within 35\mrad\ in polar angle and 50\mrad\ in azimuth of the
projected photon position on the EMC.
The second track of a \mumu pair, if within the acceptance of the calorimeter, must be
consistent with being a minimum ionizing particle.
Two identified electron or muon candidates are required for 
$\jpsi$ or $\psitwos \to \ell^+\ell^-$ reconstruction in the higher-background
$\psitwos \KS$ and $\jpsi \KL$ channels.
\par
We require a $\jpsi$ candidate to have
$2.95 \le m_{\epem} \le 3.14\gevcc$ or $3.06 \le m_{\mumu} \le 3.14\gevcc$, and a
$\psitwos\to \ell^+\ell^-$ candidate to have
$3.44 \le m_{\epem} \le 3.74\gevcc$ or $3.64 \le m_{\mumu} \le 3.74\gevcc$. 
Requirements are made on the lepton helicity angle in order
to provide further discrimination against background.
For the $\psitwos \to \jpsi \pi^+ \pi^-$ mode, mass-constrained \jpsi\ candidates
are combined with pairs of oppositely-charged
tracks considered as pions; the resulting mass must be within 
15\mevcc of the $\psitwos$ mass~\cite{PDG2000}.
\par
A $\KS\to\pi^+\pi^-$ candidate must satisfy
$489<m_{\pi^+\pi^-}<507$\mevcc.
The distance between the $\jpsi$ or \psitwos and \KS vertices is required to be at least 1\mm.
\par
Pairs of $\pi^0$ candidates with total energy above 800\mev
are considered as \KS\ candidates for the $\jpsi \KS$ mode.
We determine the most probable \KS\ decay point along the path defined
by the initial \KS\ momentum vector and the \jpsi\ vertex by maximizing
the product of probabilities for the daughter \piz\ mass-constrained fits. Allowing for vertex resolution,
we require the displacement from the \jpsi\ vertex to the decay point 
to be between $-10$ and $+40$\cm\ and
the $\piz\piz$ mass evaluated at this point to be between 470 and 550\mevcc.
\par
A \KL candidate is formed from a cluster not matched to a reconstructed track. For the
EMC the cluster must have energy above 200\mev, while for the IFR the cluster must have 
at least two layers.
We determine the \KL energy by combining its direction with the reconstructed 
$\jpsi$ momentum, assuming the decay $\Bz \rightarrow \jpsi \KL$. 
To reduce photon backgrounds, EMC clusters consistent with a $\piz\to\gamma\gamma$ decay
are rejected and the transverse missing momentum of the event projected on
the \KL candidate direction must be
consistent with the \KL\ momentum. In addition,
the center-of-mass \jpsi\ momentum is required to be greater than 1.4\gevc.

\begin{figure}[!]
\begin{center}
\epsfxsize8.6cm
\figurebox{}{}{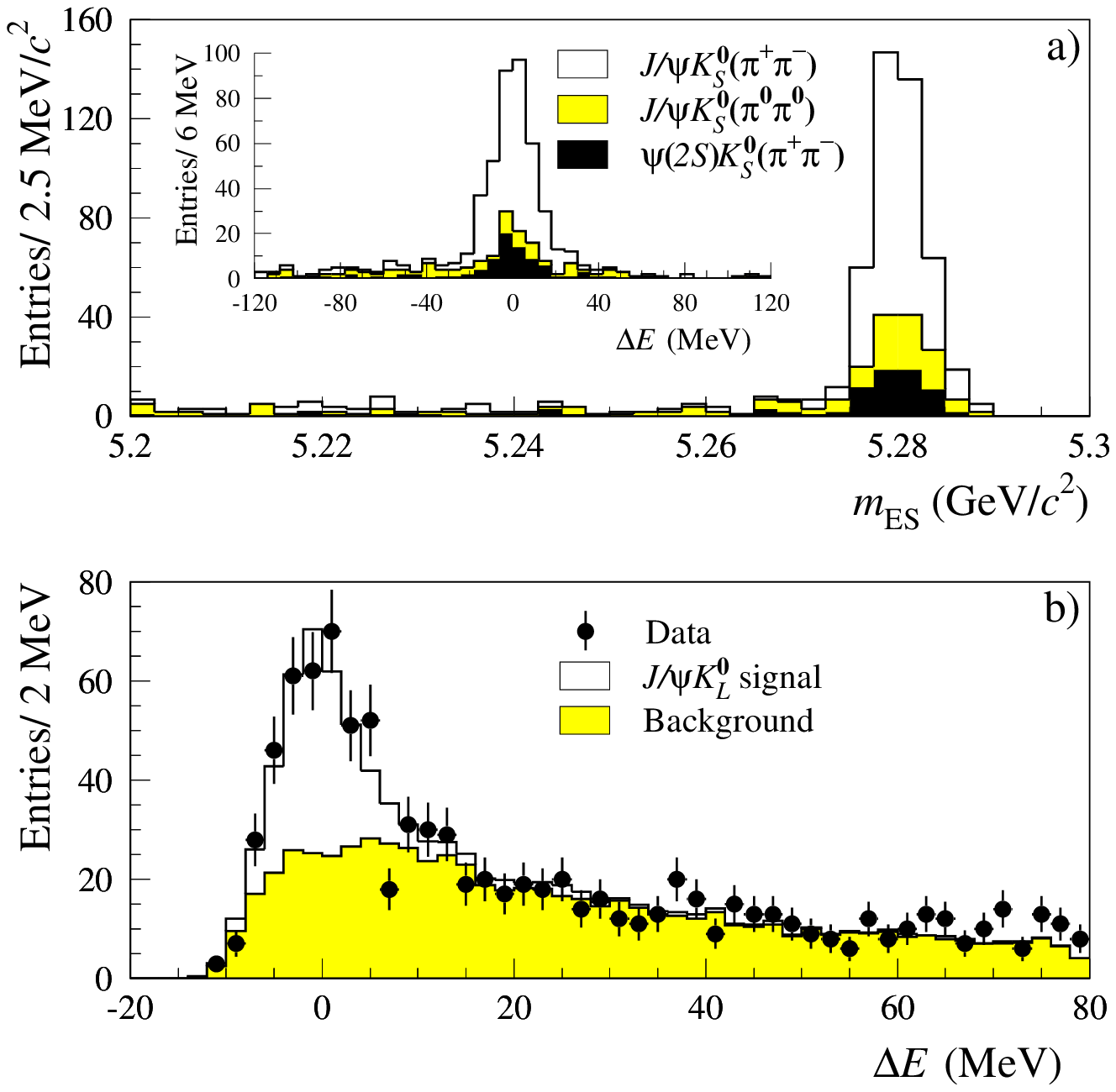}
\caption{ a) Distribution of \mes\ and $\Delta E$ for $B_{\CP}$ candidates having a \KS
in the final state;
b) distribution of $\Delta E$ for $\jpsi\KL$ candidates.}
\label{fig:prlfig1}

\end{center}
\end{figure}

\par
$B_{CP}$ candidates used in the analysis are selected by
requiring that the difference $\Delta E$ between the energy of the $B_{CP}$ candidate 
and the beam energy in the center-of-mass frame be less than three standard deviations
from zero and that, for \KS\ modes,
the beam-energy substituted mass $\mes=\sqrt{{(E^{\rm cm}_{\rm beam})^2}-(p_B^{\rm cm})^2}$ 
falls in the interval
$\mes >  5.2\gevcc$. The resolution for $\Delta E$ is about 10\mev, except for $\jpsi\KL$ (3\mev)
and the $\KS\to\piz\piz$ mode (33\mev).
For the purpose of determining numbers of events, purities and efficiencies,
a signal region $\mes > 5.27\gevcc$ is used for
all modes except $\jpsi\KL$.
\par
Figure~\ref{fig:prlfig1} shows the resulting $\Delta E$ and \mes distributions
for $B_{\CP}$ candidates containing a \KS, and $\Delta E$ for the 
candidates containing a \KL.
The $B_{\CP}$ sample is composed of 890 events in the signal region, 
with an estimated background of 260 events, 
predominantly in the $\jpsi\KL$ channel. 
For that channel, the
composition, effective $\eta_f$ and $\Delta E$ distributions of the individual background sources
are taken either from a Monte Carlo simulation (for $B$ decays to \jpsi) or
the $m_{\ell^+ \ell^-}$ sidebands in data.
\par
For flavor tagging, we exploit information from the incompletely 
reconstructed other $B$ decay in the event. 
The charge of energetic electrons and muons from 
semileptonic $B$ decays, kaons, soft pions from \Dstar decays, and  
high momentum charged particles is correlated with the flavor of the decaying $b$ quark: 
{\it e.g.}, a positive lepton yields a \Bz tag.
Each event is assigned to one of four hierarchical,
mutually exclusive 
tagging categories or is excluded from further analysis. 
The mistag fractions and efficiencies of all categories are determined from data.
\par 
A lepton tag requires an electron or muon candidate with a 
center-of-mass momentum $p_{\rm cm} >1.0$ or $1.1\gevc$, respectively. This efficiently selects
primary leptons and
reduces contamination due to oppositely-charged leptons from semileptonic 
charm decays.
Events meeting these criteria are assigned to the {\tt Lepton} 
category unless the lepton charge and the net charge of all kaon candidates indicate opposite tags. 
Events without a lepton tag but with a non-zero net kaon charge are assigned to the 
{\tt Kaon} category. 
\par
All remaining
events are passed to a neural network algorithm 
whose main inputs are the momentum and charge 
of the track with the maximum center-of-mass 
momentum, and the outputs of secondary networks,
trained with Monte Carlo samples to identify primary leptons, kaons, 
and soft pions. 
Based on the output of the neural network algorithm, events are tagged as \Bz or \Bzb and 
assigned to the {\tt NT1} (more certain tags) or {\tt NT2} (less certain tags) category, or 
not tagged at all. 
The tagging power of the {\tt NT1} and {\tt NT2} categories arises primarily 
from soft pions  
and from recovering unidentified isolated primary electrons and muons. 

Table~\ref{tab:result} shows the number of tagged events and the signal purity, determined
from fits to the \mes (\KS\ modes) or $\Delta E$ (\KL\ mode) distributions. 
The measured efficiencies for the four tagging categories are summarized in 
Table~\ref{tab:TagMix:mistag}.

\begin{table}[!htb] 
\caption{ 
Number of tagged events, signal purity and result of fitting for \CP\ asymmetries in 
the full \CP sample and in 
various subsamples, as well as in the $B_{\rm flav}$ and charged $B$ control samples.  
Purity is the fitted number of signal events divided by the total number of 
events in the $\Delta E$ and \mes signal region defined in the text. Errors are statistical only.}
\label{tab:result} 
\begin{ruledtabular} 
\begin{tabular*}{\hsize}{ l@{\extracolsep{0ptplus1fil}} r c@{\extracolsep{0ptplus1fil}} D{,}{\ \pm\ }{-1} } 
 Sample                                  & $N_{\rm tag}$    & Purity (\%)    &  \multicolumn{1}{c}{$\ \ \ \stwob$}  \\ \colrule 
$\jpsi\KS$, $\psitwos\KS$                & $273$        & $96\pm1$       &  0.25,0.22   \\ 
$\jpsi \KL$                              & $256$        & $39\pm6$       &  0.87,0.51   \\ 
\hline
 Full \CP\ sample                        & $529$        & $69\pm2$       &  0.34,0.20   \\ 
\hline
\hline
$\jpsi\KS$, $\psitwos\KS$ only & & & \\
\hline
$\ \jpsi \KS$ ($\KS \to \pi^+ \pi^-$)    & $188$        & $98\pm1$       &  0.25,0.26   \\ 
$\ \jpsi \KS$ ($\KS \to \pi^0 \pi^0$)    & $41$         & $85\pm6$       &  -0.05,0.66  \\ 
$\ \psi(2S) \KS$ ($\KS \to \pi^+ \pi^-$) & $44$         & $97\pm3$       &  0.40,0.50   \\ 
\hline 
$\ $ {\tt Lepton} tags                   & $34$         &  $99\pm2$      &  0.07,0.43   \\ 
$\ $ {\tt Kaon} tags                     & $156$        &  $96\pm2$      &  0.40,0.29    \\ 
$\ $ {\tt NT1} tags                      & $28$         &  $97\pm3$      &  -0.03,0.67    \\ 
$\ $ {\tt NT2} tags                      & $55$         &  $96\pm3$      &  0.09,0.76    \\ 
\hline 
$\ $ \Bz\ tags                           & $141$        &  $96\pm2$      &  0.24,0.31     \\ 
$\ $ \Bzb\ tags                          & $132$        &  $97\pm2$      &  0.25,0.30     \\ 
\hline\hline
$B_{\rm flav}$ sample                    & $4637$       & $86\pm1$       &  0.03,0.05     \\
\hline 
Charged $B$ sample                       & $5165$       & $90\pm1$       &  0.02,0.05     \\

\end{tabular*} 
\end{ruledtabular} 
\end{table}

\begin{table}[!htb] 
\caption
{ Average mistag fractions $\mistag_i$ and mistag differences $\Delta\mistag_i=\mistag_i(\Bz)-\mistag_i(\Bzb)$
extracted for each tagging category $i$ from the maximum-likelihood fit to the time distribution for the 
fully-reconstructed \Bz\ sample ($B_{\rm flav}$+$B_{\CP}$). The figure of merit for tagging is 
the effective tagging efficiency $Q_i = \eps_i (1-2\mistag_i)^2$, where $\eps_i$ 
is the fraction of events with a reconstructed tag vertex that are
assigned to the $i^{th}$ category. 
Uncertainties are statistical only. The 
statistical error on \stwob is proportional to $1/\sqrt{Q}$, where $Q=\sum Q_i$. } 
\label{tab:TagMix:mistag} 
\begin{ruledtabular} 
\begin{tabular*}{\hsize}{ l@{\extracolsep{0ptplus1fil}} D{,}{\ \pm\ }{-1} @{\extracolsep{10ptplus1fil}} c@{\extracolsep{0ptplus1fil}} D{,}{\ \pm\ }{-1} @{\extracolsep{0ptplus1fil}} D{,}{\ \pm\ }{-1} } 
Category     & \multicolumn{1}{c}{$\ \ \ \varepsilon$ (\%)} & $\mistag$ (\%) & \multicolumn{1}{c}{$\ \ \ \ \Delta\mistag$ (\%)} & \multicolumn{1}{c}{$\ \ \ Q$ (\%)}       \\ \colrule 
{\tt Lepton} & 10.9,0.4 & $11.6\pm2.0$ & 3.1,3.1  &   6.4,0.7  \\ 
{\tt Kaon}   & 36.5,0.7 & $17.1\pm1.3$ & -1.9,1.9 &  15.8,1.3  \\ 
{\tt NT1}    &  7.7,0.4 & $21.2\pm2.9$ & 7.8,4.2  &   2.6,0.5  \\ 
{\tt NT2}    & 13.7,0.5 & $31.7\pm2.6$ & -4.7,3.5 &   1.8,0.5  \\  \colrule 
All          & 68.9,1.0 &              &          &  26.7,1.6  \\ 
\end{tabular*} 
\end{ruledtabular} 
\end{table} 

\par
The uncertainty in the \deltat measurement is
dominated by the measurement of the position $z_{\rm tag}$ of the tagging vertex.
The tagging vertex is determined by fitting the
tracks not belonging to
the $B_{\CP}$ (or $B_{\rm flav}$) candidate to a common vertex. 
Reconstructed \KS and $\Lambda$ candidates are used as input to the fit in place of their 
daughters. Tracks from $\gamma$ conversions
are excluded from the fit. To reduce contributions from charm decay, which
bias the vertex estimation, 
the track with the largest vertex $\chi^2$ contribution
greater than 6 is removed and the fit is redone
until no track fails the $\chi^2$ requirement or fewer than two tracks remain.
The average resolution for $\deltaz=z_{CP}-z_{\rm tag}$ is 190\mum.
The time interval \deltat between the two $B$ decays is then determined
from the \deltaz measurement, including an event-by-event correction for the direction
of the $B$ with respect to the $z$ direction in the $\FourS$ frame. 
An accepted candidate must have a converged fit for the $B_{CP}$ and $B_{\rm tag}$ vertices,
an error of less than 400\mum on \deltaz\ and a measured $\vert \deltaz \vert < 3\mm$;
86\% of the $B_{CP}$ events satisfy this requirement.
\par
The \stwob measurement is made with an unbinned maximum likelihood fit to the \deltat 
distribution of the combined $B_{CP}$ and $B_{\rm flav}$ tagged samples.
The \deltat\ distribution of the former is given by Eq.~\ref{eq:timedist}, with $|\lambda|=1$. The latter
evolves according to the known rate
for flavor oscillations in neutral $B$ mesons. The amplitudes for $B_{CP}$ asymmetries
and for $B_{\rm flav}$ flavor oscillations are reduced
by the same factor $(1-2\mistag)$ due to mistags. The distributions are both convoluted with a common
\deltat resolution function, and are corrected for backgrounds, incorporated with different assumptions about their
\deltat evolution and convoluted with a
separate resolution function.
Events are assigned signal and background probabilities based on fits to 
\mes\ (all modes except $\jpsi\KL$) or $\Delta E$ ($\jpsi\KL$) distributions. 
\par
The \deltat\ resolution function for signal candidates is represented
by a sum of three Gaussian distributions with different means and
widths. For the core and tail Gaussians, the widths
are scaled by the event-by-event measurement error
derived from the vertex fits; the combined rms is 1.1\ps. A separate offset for the core distribution
is allowed for each tagging category to account for small shifts
caused by inclusion of residual charm decay products in the tag vertex;
a common offset is used for the tail component.  
The third Gaussian (of fixed 8\ps\ width)
accounts for the fewer than $1\%$ of events with incorrectly
reconstructed vertices. 
Identical resolution function parameters are used for all
modes, since the $B_{\rm tag}$ vertex precision dominates
the \deltat resolution. 
\par
A total of 35 parameters are varied in the final fit, including
the values of \stwob (1), the average mistag fraction $\mistag$ and the 
difference $\Delta\mistag$ between \Bz\ and \Bzb\ mistags for each tagging category (8), parameters for the signal \deltat
resolution (9), and parameters for background time dependence (6), \deltat resolution (3) and mistag fractions (8).
The determination of the mistag fractions and signal \deltat resolution function is dominated
by the high-statistics $B_{\rm flav}$ sample, while background parameters are governed
by events with $\mes < 5.27\gevcc$ (except $\jpsi\KL$). 
We fix $\tau_{\Bz}=1.548\ps$ and 
$\Delta m_{B^0}=0.472\,\hbar\ps^{-1}$~\cite{PDG2000}.
The largest correlation between \stwob\ and any linear combination of the other free
parameters is 0.076.

\begin{figure}[t!]
\begin{center}
\epsfxsize8.6cm
\figurebox{}{}{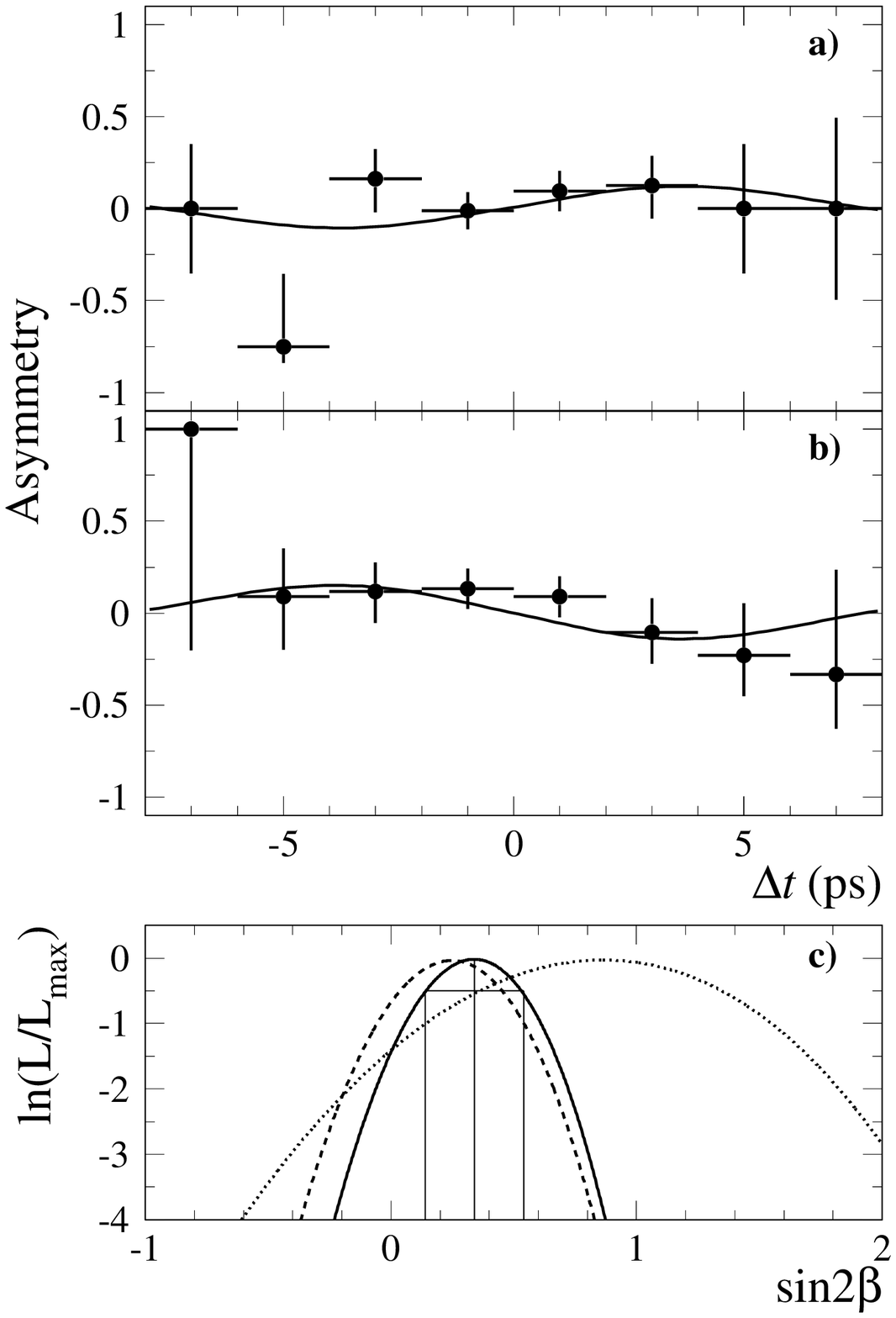}
\caption{ The raw 
asymmetry in the number of \Bz\ and \Bzb\ tags in the signal region,
$(N_{\Bz}-N_{\Bzb})/(N_{\Bz}+N_{\Bzb})$, with asymmetric binomial errors,
as a function of \deltat for a) the $\jpsi\KS$ and $\psitwos\KS$ modes $(\eta_f=-1)$ 
and b) the $\jpsi\KL$ mode $(\eta_f=+1)$.
The solid curves represent the time-dependent asymmetries determined for the central values of \stwob 
from the fits for these samples. Eight events that lie outside the plotted interval were 
also used in the fits.
The probability of obtaining a lower likelihood, evaluated using a Monte Carlo technique, is 60\%.
c) Variation of the log likelihood as a 
function of \stwob for the modes  containing \KS (dashed curve), the
$\jpsi\KL$ mode (dotted) and the entire sample (solid).
For the latter, solid lines indicate the central value and values of the log likelihood
corresponding to one statistical standard deviation.}
\label{fig:asymlike}
\end{center}
\end{figure}

\par
The measurement of \stwob\ was performed as a blind analysis
by hiding the value of \stwob obtained from the fit, as well as the \CP\ asymmetry
in the \deltat\ distribution, until the analysis was complete.
This allowed us to study statistical and systematic errors 
without knowing the numerical value of \stwob.
\par
The measured mistag rates obtained from the likelihood fit 
for the four tagging categories are summarized in 
Table~\ref{tab:TagMix:mistag}.
As a check, the mistag rates were evaluated
with a sample of about 16,000 $D^{*-} \ell^+ \nu_\ell$ events 
and found to be consistent with the results from the hadronic decay sample. 
\par
The combined fit to the \CP decay modes and the flavor decay modes yields
\begin{eqnarray}
\stwob=0.34 \pm 0.20\ \stat \pm 0.05\ \syst. \nonumber
\end{eqnarray}
The decay asymmetry
${A}_{\CP}$ as a function of \deltat and the log likelihood as a function of \stwob 
are shown in Fig.~\ref{fig:asymlike}. 
If $\vert\lambda\vert$ is allowed to float in the fit, the value obtained is consistent with 1 and there
is no significant difference in the value of $-\eta_f\mathop{\cal I\mkern -2.0mu\mit m}
\lambda/|\lambda|$ (identified with \stwob\ in the Standard Model) and our quoted result.
Repeating the fit with all parameters fixed to their determined values except \stwob, we find
a total contribution of $\pm 0.02$ to the error on \stwob\ is due to the combined statistical
uncertainties in mistag rates, \deltat\ resolution and background parameters.
\par
The dominant sources of systematic error are the assumed parameterization of 
the \deltat\ resolution function (0.04),
due in part to residual uncertainties in the SVT alignment,  
and uncertainties in the level, composition, and \CP\ asymmetry of the background
in the selected \CP events (0.02). 
The systematic errors from
uncertainties in $\Delta m_{\Bz}$ and $\tau_{\Bz}$ and from the
parameterization of the background
in the selected $B_{\rm flav}$ sample are found to be negligible. 
An increase of $0.02\,\hbar\ps^{-1}$ in the assumed value for $\Delta m_{\Bz}$ 
decreases \stwob\ by 0.012.
\par
The large sample of reconstructed events allows a number of consistency
checks, including separation of the data by decay mode, tagging
category and $B_{\rm tag}$ flavor. The results of fits to these subsamples
are shown in Table~\ref{tab:result} for the high-purity \KS\ events. Table~\ref{tab:result} also
shows results of fits with the samples of non-\CP decay modes, where no statistically
significant asymmetry is found.
\par
Our measurement of \stwob is consistent with, but improves substantially on the precision of, previous
determinations~\cite{OPALCDFALEPH}. The central value is consistent with the range implied
by measurements and theoretical estimates of the magnitudes of CKM matrix elements~\cite{CKMconstraints};
it is also consistent with no \CP asymmetry at the 1.7$\sigma$ level.
\par
We wish to thank our \pep2\ colleagues for their extraordinary
achievement in reaching design luminosity and high reliability in
a remarkably short time. The collaborating institutions wish to thank 
SLAC for its support and the kind hospitality extended to them.   
\par
This work has been supported by the US Department of Energy and National Science
Foundation, the Natural Sciences and Engineering
Research Council (Canada), the Institute of High Energy Physics (China), Commissariat
\`a l'Energie Atomique
and Institut National de Physique Nucl{\'e}aire et de Physique des Particules (France),
Bundesministerium
f{\"u}r Bildung und Forschung (Germany), Istituto Nazionale di Fisica Nucleare (Italy),
the Research
Council of Norway, the Ministry of Science and Technology of the Russian Federation, and the
Particle Physics
and Astronomy Research Council (United Kingdom). Individuals have 
received support from the Swiss National Foundation, the A. P. Sloan Foundation, 
the Research Corporation
and the Alexander von Humboldt Foundation.
  
\vskip12pt 
\hbox{$\ \ \ \ \ \ \ \ \ \ \ \ \ \ \ \ \ \ \ \ \ \ \ \ ${\LARGE{\bf---------------}}}

\end{document}

%% file: authors.tex
\author{B.~Aubert}
\author{D.~Boutigny}
\author{I.~De Bonis}
\author{J.-M.~Gaillard}
\author{A.~Jeremie}
\author{Y.~Karyotakis}
\author{J.P.~Lees}
\author{P.~Robbe}
\author{V.~Tisserand}
\affiliation{Laboratoire de Physique des Particules, F-74941 Annecy-le-Vieux, France }
\author{A.~Palano}
\affiliation{Universit\`a di Bari, Dipartimento di Fisica and INFN, I-70126 Bari, Italy }
% ...
